\documentclass[
aip,rsi,
amsmath,amssymb,
superscriptaddress,
reprint]{revtex4-1}

\usepackage[T1]{fontenc}
\usepackage[utf8]{inputenc}

\usepackage{graphicx}
\usepackage[hidelinks]{hyperref}

\begin{document}

\title{Intermodulation spectroscopy as an alternative to pump-probe for the measurement of fast dynamics at the nanometer scale}

\author{Riccardo Borgani}
\email{borgani@kth.se}
\affiliation{Nanostructure Physics, KTH Royal Institute of Technology, 10691 Stockholm, Sweden}

\author{David B. Haviland}
\affiliation{Nanostructure Physics, KTH Royal Institute of Technology, 10691 Stockholm, Sweden}

\date{\today}

\begin{abstract}
We present an alternative approach to pump-probe spectroscopy for measuring fast charge dynamics with an atomic force microscope (AFM).
Our approach is based on coherent multifrequency lock-in measurement of the intermodulation between a mechanical drive and an optical or electrical excitation.
In response to the excitation, the charge dynamics of the sample is reconstructed by fitting a theoretical model to the measured frequency spectrum of the electrostatic force near resonance of the AFM cantilever.
We discuss the time resolution, which in theory is limited only by the measurement time, but in practice is of order one nanosecond for standard cantilevers and imaging speeds.
We verify the method with simulations and demonstrate it with a control experiment, achieving a time resolution of $20~\mathrm{ns}$ in ambient conditions, limited by thermal noise.

\end{abstract}

\maketitle

\section{Introduction}
Characterizing fast dynamical processes at the nanometer scale is key to understanding and optimizing the relation between structure and function in nanotechnology, where a particularly active field is the study of photo-induced charge dynamics in energy materials\cite{Rao2013,Weber2018}.
Optical pump-probe experiments\cite{Cerullo2007} routinely achieve femtosecond time resolution\cite{Polli2010}, with progress towards the attosecond regime\cite{Calegari2014}, but they are typically diffraction-limited to hundreds of nanometers spatial resolution.
Atomic force microscopy (AFM) is an ideal method for investigating material properties at the nanometer scale and considerable effort has been put into pushing the limits of time resolution in AFM.
Techniques inspired by optical pump-probe spectroscopy require multiple measurements at each location\cite{Weber2018}, with different pulse rates\cite{FernandezGarrillo2016} or different pump-probe delays\cite{Schumacher2017}.
Others require handling large datasets\cite{Collins2017} or advanced filtering routines\cite{Karatay2016}.
In this manuscript we demonstrate coherent multifrequency AFM methods that capture fast dynamics through analysis of several closely-spaced Fourier components of the force, all measured near the cantilever resonance where sensitivity is greatest.

Pump-probe spectroscopy explores fast dynamics with two pulses and a fixed delay.
The response of the sample~$R$ is measured while a pump pulse excites some fast process, rapidly followed by a second probe pulse at fixed delay time~$t^\prime$.
Repeating this measurement at different delay times gives~$R(t^\prime)$, which contains information about the fast dynamics.
The delay can be as short as the pulse width, and because it can be kept constant, long-time averaging over many identical events gives the desired signal-to-noise ratio (SNR).
The events typically have some repetition frequency, but stability of this frequency is not a stringent requirement as each pump-probe event is considered statistically independent and the assumption is that the sample relaxes to the same initial state in between events.

Here we present frequency-domain alternatives to pump-probe which are also capable of resolving fast dynamics, in spite of the limited bandwidth inherent to a sensitive detector.
The frequency domain approach exploits periodic signals or pulse trains that are carefully tuned such that they are coherent with one another, \textit{i.e.} they have a fixed phase relation to a single reference oscillation.
This allows for lock-in measurement of the amplitude and phase of many intermodulation products generated by the nonlinear detection process.
The fast time-domain response is then reconstructed through Fourier analysis of the measured intermodulation spectrum.

Intermodulation spectral methods do not rely on the constancy of the delay.
Rather, they exploit the stability of a reference oscillation, something that can be achieved with great precision\cite{Ludlow2015}.
Because the pump and probe signals are tuned, the time-domain delay between these signals changes in a regular manner over the entire long-time integration of the response signal, needed to determine the Fourier coefficients.
At first sight this frequency-domain approach may appear more complex, but it comes with an advantage over pump-probe:
tuned multifrequency lock-in measurement gives coherent signal averaging, where all frequencies in the measured spectrum are demodulated in parallel, during the same time window.
This frequency-domain multiplexing reduces the measurement time needed to resolve the fast dynamics at the desired SNR.

In the following we describe in detail the principles of intermodulation spectroscopy.
We derive the theoretical limit of achievable time resolution for several different intermodulation methods, and we verify them through simulation and experiment.
These different methods differ in their excitation schemes, but they all have the common feature that the material response is probed by a measurement of the force spectrum near the cantilever resonance, where force measurement sensitivity is at the thermal limit.

\section{Measuring force with dynamic AFM}

At frequencies near its first flexural eigenmode, the AFM cantilever is well approximated by a driven dampened harmonic oscillator described by the ordinary differential equation:

\begin{equation}
    \ddot{d} + \frac{\omega_0}{Q}\dot{d}+\omega_0^2d=\frac{\omega_0^2}{k}F,
    \label{eq:ode}
\end{equation}
where $d(t)$ is the deflection of the cantilever from its equilibrium position, and $F$ is the total force acting on it.
The calibration constants $\omega_0$, $Q$ and $k$ are the mode resonance frequency, quality factor and stiffness, respectively.
It is convenient to work with the Fourier transform of Eq.~(\ref{eq:ode}):
\begin{equation}
    \hat{d}(\omega) = \hat{\chi}(\omega)\hat{F}(\omega),
\end{equation}
where $\hat{\chi}(\omega)$ is the linear response function of the cantilever
\begin{equation}
    \hat{\chi}(\omega) = \frac{1}{k}\left(1-\frac{\omega^2}{\omega_0^2} + \mathrm{i}\frac{\omega}{\omega_0 Q}\right)^{-1}.
    \label{eq:chi}
\end{equation}
In our notation, we use~$\hat{f}(\omega)$ to denote a complex-valued function of frequency given by the Fourier transform of the real-valued function of time~$f(t)$.
In the following we often drop the explicit frequency and time dependence.

When the AFM probe is lifted far away from the sample surface and the cantilever is in thermal equilibrium with the surrounding damping medium, a measurement of the power spectral density in a narrow band around the high-$Q$ resonance reveals the thermal fluctuations of cantilever deflection, above the detector noise floor.
From this measurement we obtain the calibration constants $\omega_0$, $Q$ and $k$, as well as the responsivity $\alpha$ ($\mathrm{V}/\mathrm{nm}$) of the deflection detector \cite{Sader2012,Higgins2006,Sader2016}.

Driving the cantilever with a force $F_\mathrm{drive} = F_\mathrm{D} \cos (\omega_\mathrm{D} t + \psi_\mathrm{D})$ at a frequency~$\omega_\mathrm{D} \approx \omega_0$ results in the ``free'' motion
\begin{equation}
    \hat{d}_\mathrm{free} = \hat{\chi} \hat{F}_\mathrm{drive}.
\end{equation}
Measuring $\hat{d}_\mathrm{free}$ provides knowledge of the drive force, eliminating the need for an independent calibration of the actuator.
As the probe gets closer to the surface, additional linear forces act on the body of the cantilever which must be accounted for.
These interactions are called background forces, and we compensate for their effect on the measured data by determining their linear response function~$\hat{\chi}_\mathrm{BG}$, using a technique described in a previous publication\cite{Borgani2017}.

With the probe at the sample surface, we obtain the nonlinear ``tip-sample'' force from a measurement of the cantilever deflection $\hat{d}$:
\begin{equation}
    \hat{F}_\mathrm{TS} = \hat{\chi}^{-1}\left(\hat{d} - \hat{d}_\mathrm{free}\right) - \hat{\chi}^{-1}_\mathrm{BG}\hat{d} .
    \label{eq:Fts_from_d}
\end{equation}

\section{Electrostatic force from cantilever dynamics}
\begin{figure*}
    \centering
    \includegraphics[width=\textwidth]{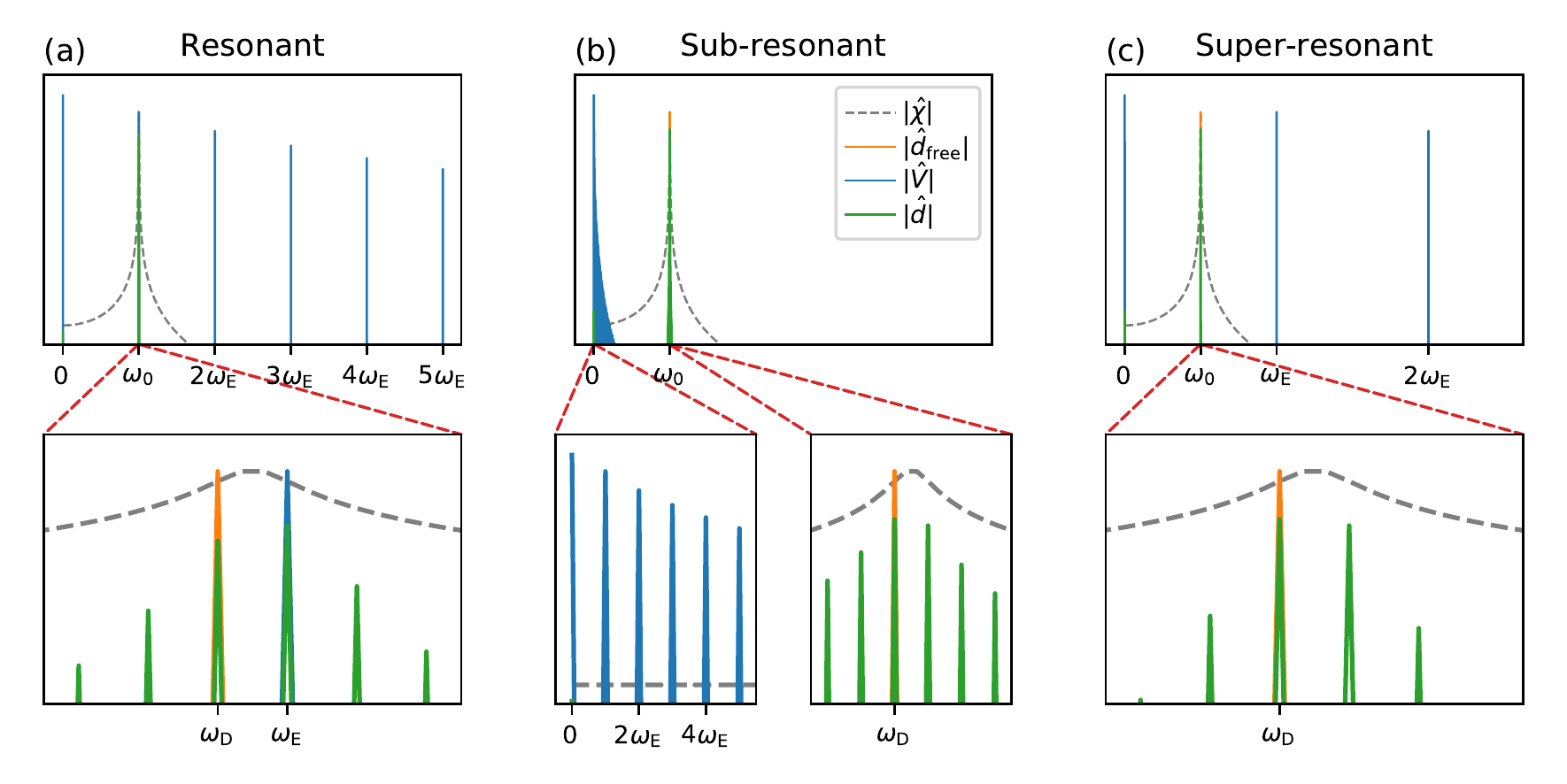}
    \caption{Frequency-domain representation of the three proposed excitation schemes, with cantilever drive frequency $\omega_\mathrm{D} \approx \omega_0 \gg \delta$.
    (a) Resonant: the electrical response is excited at $\omega_\mathrm{E} = \omega_\mathrm{D} + \delta$.
    (b) Sub-resonant: the electrical response it excited at low frequency $\omega_\mathrm{E} = \delta$.
    (c) Super-resonant: the electrical response is excited close to the second harmonic of $\omega_\mathrm{D}$, at frequency $\omega_\mathrm{E} = 2\omega_\mathrm{D} + \delta$.
    In all cases, intermodulation products arise close to resonance, equally spaced by $\delta$.
    }
    \label{fig:freq}
\end{figure*}
\begin{figure*}
    \centering
    \includegraphics[width=\textwidth]{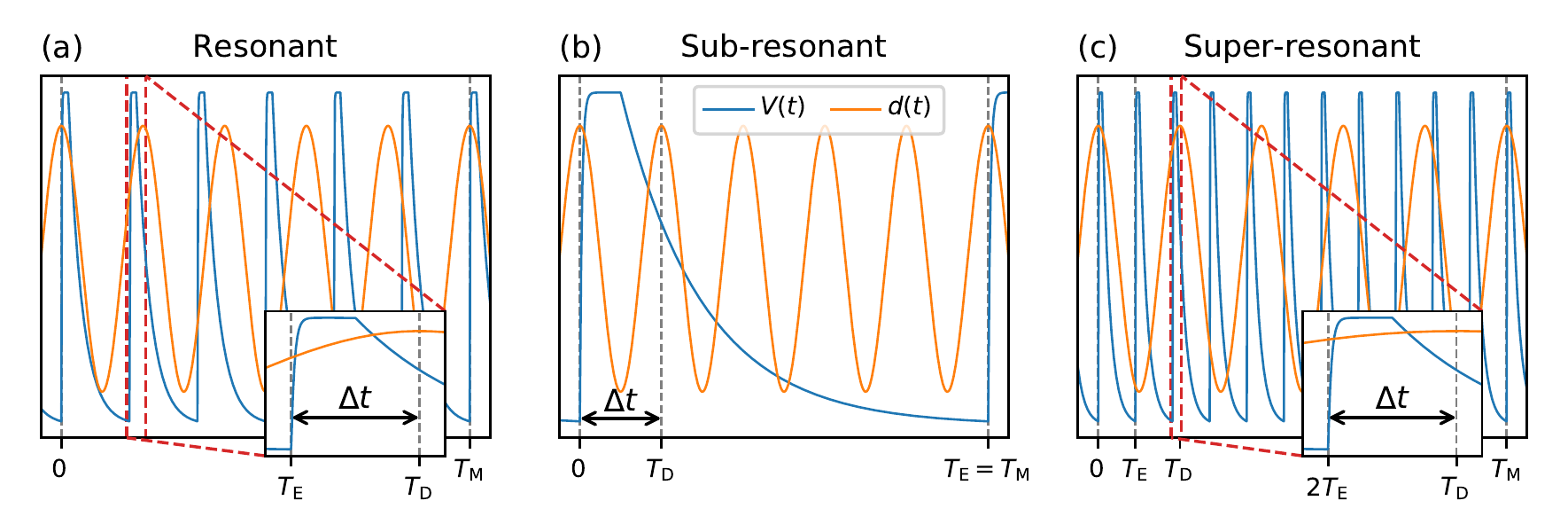}
    \caption{Time-domain representation of the three proposed excitation schemes, for fixed measurement time $T_\mathrm{M} = 2\pi / \delta$ and cantilever drive period $T_\mathrm{D} = 2\pi/\omega_\mathrm{D} \ll T_\mathrm{M}$.
    (a) Resonant: the electrical response is excited with period $T_\mathrm{E} = 2\pi/\omega_\mathrm{E} = 2\pi/(\omega_\mathrm{D} + \delta)$, the time resolution is set by the minimum time delay between the electrical excitation and the cantilever oscillation $\Delta t = T_\mathrm{D}-T_\mathrm{E}\approx2\pi\delta / \omega_0^2$.
    (b) Sub-resonant: the electrical response it excited once in the measurement window with period $T_\mathrm{E} = T_\mathrm{M}$, the time resolution is given by the single oscillation cycle of the cantilever $\Delta t = T_\mathrm{D} \approx 2\pi / \omega_0$.
    (c) Super-resonant: the electrical response is excited close to the second harmonic of $\omega_\mathrm{D}$, with period $T_\mathrm{E} = 2\pi/(2\omega_\mathrm{D} + \delta)$.
    The minimum time delay is $\Delta t = T_\mathrm{D}-2T_\mathrm{E}\approx\pi\delta / \omega_0^2$.
    The curves shown are a conceptual sketch, in reality the difference between $T_\mathrm{D}$ and $T_\mathrm{M}$ is much bigger, \textit{i.e.} there are many more cantilever oscillations (about $Q$) in one measurement window.
    }
    \label{fig:res}
\end{figure*}
\begin{figure}
    \centering
    \includegraphics[width=\columnwidth]{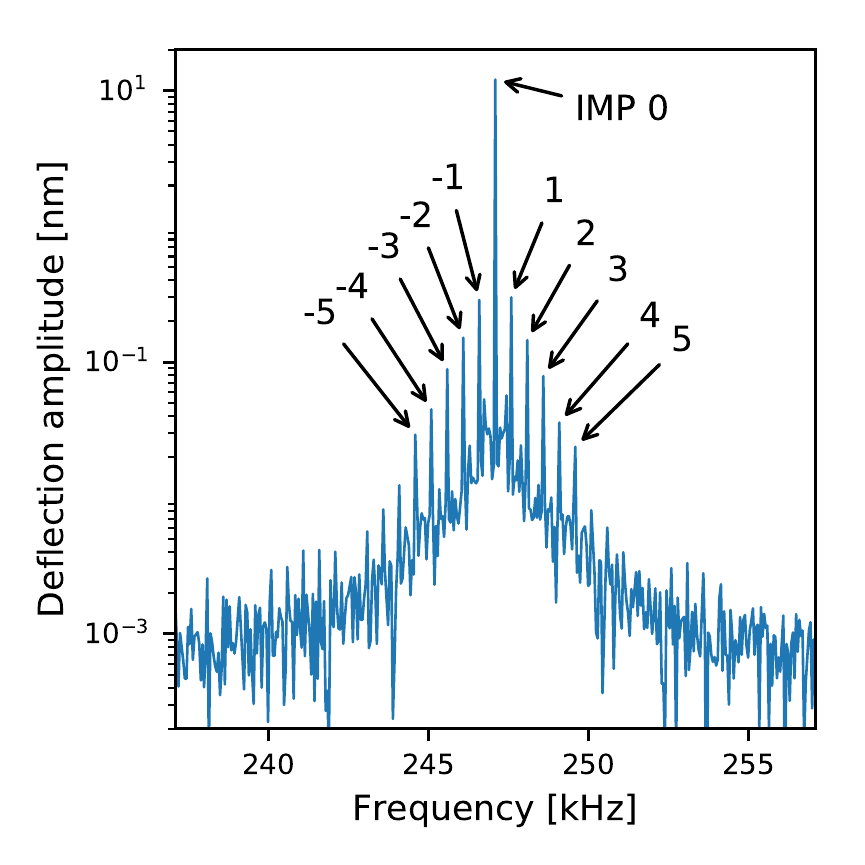}
    \caption{Spectrum of the cantilever deflection near resonance.
    The intermodulation products (IMPs) are numbered according to their position with respect to the mechanical drive frequency $\omega_\mathrm{D}$.
    The plotted data are from a control experiment with sub-resonant excitation, $\tau_\mathrm{F} = 200~\mathrm{\mu s}$, and $\delta=2\pi\times 500~\mathrm{Hz}$.
    }
    \label{fig:imps}
\end{figure}

In electrostatic force microscopy (EFM), the dominant contribution to the tip-surface force is a nonlinear function of the potential difference $V$ between the tip and the sample,
\begin{equation}
    F_\mathrm{TS} = \frac{1}{2}\frac{\partial C}{\partial z} V^2,
    \label{eq:Fts}
\end{equation}
where the sample surface is at $z=0$ so that $z(t) = d(t) + h$ is the instantaneous tip-sample separation when the cantilever is deflected by~$d$ from the rest position~$h$.
The capacitance gradient~$\partial C/\partial z$ is generally a nonlinear function of~$z$.

Because the mechanical driving force $F_\mathrm{drive}$ is much larger than the perturbing electrostatic forces, the tip-sample separation to first-order in perturbation theory is
\begin{equation}
    z(t) \approx A_\mathrm{D} \cos(\omega_\mathrm{D} t + \psi_\mathrm{D}) + h.
    \label{eq:zt}
\end{equation}
If the capacitance gradient is assumed to be analytic, it has the same time periodicity as~$z$.
We may therefore expand it in a Fourier series of~$\omega_\mathrm{D}$
\begin{equation}
    \frac{1}{2}\frac{\partial C}{\partial z}[z(t)] = \sum_{k=-\infty}^{+\infty} c_k \mathrm{e}^{\mathrm{i}(k\omega_\mathrm{D} t + \phi_k)},
    \label{eq:Czt}
\end{equation}
where~$\{c_k\}$ and~$\{\phi_k\}$ are real numbers such that~$c_{-k}=c_k$ and~$\phi_{-k}=-\phi_k$.
The coefficient~$c_k \mathrm{e}^{\mathrm{i}\phi_k}$ is the complex amplitude of the discrete Fourier transform (DFT) of~${\partial C}/{\partial z}$ at frequency~$k\omega_\mathrm{D}$.
Note that while the first-order motion has only one component at frequency~$\omega_\mathrm{D}$, the capacitance gradient has components at all harmonics of~$\omega_\mathrm{D}$, due to its nonlinearity.

We excite the sample with a periodic train of pulses (optical or electrical) with repetition frequency~$\omega_\mathrm{E}$.
In response to these pulses a surface potential is generated and we may, similarly, expand the nonlinear term~$V^2$ in a Fourier series of~$\omega_\mathrm{E}$:
\begin{equation}
    V^2(t) = \sum_{j=-\infty}^{+\infty} v_j \mathrm{e}^{\mathrm{i}(j\omega_\mathrm{E} t + \theta_j)},
    \label{eq:V2t}
\end{equation}
where~$\{v_j\}$ and~$\{\theta_j\}$ are real numbers such that~$v_{-j}=v_j$ and~$\theta_{-j}=-\theta_j$, and~$v_j \mathrm{e}^{\mathrm{i}\theta_j} = \widehat{V^2}(j\omega_\mathrm{E})$.

The product of Eq.s~(\ref{eq:Czt}) and~(\ref{eq:V2t}) introduces terms in Eq.~(\ref{eq:Fts}) for $\hat{F}_\mathrm{TS}$ at frequencies that are integer linear combinations of~$\omega_\mathrm{D}$ and~$\omega_\mathrm{E}$, the so-called intermodulation products (IMPs), or frequency-mixing products:
\begin{equation}
    \hat{F}_\mathrm{TS}(k\omega_\mathrm{D} + j\omega_\mathrm{E}) = c_k v_j \mathrm{e}^{\mathrm{i}(\phi_k + \theta_j)}.
    \label{eq:all_imps}
\end{equation}
Depending on the choice of frequencies of the mechanical drive and electrical excitation, some IMPs arise close to the cantilever resonance where the high quality factor allows for a measurement of deflection (and therefore force) with the highest possible SNR.
Below we analyze three cases where the optical or electrical excitation at~$\omega_\mathrm{E}$ is close to, below, and above resonance, while the mechanical drive at~$\omega_\mathrm{D}$ is kept at resonance.

\subsection{Resonant excitation}
The resonant scheme is analogous to that used in intermodulation AFM\cite{Platz2008}, and has strong similarities to the traditional optical pump-probe method described above.
The cantilever and the electrical response are both excited close to the resonance frequency at $\omega_\mathrm{D} \approx \omega_0$ and $\omega_\mathrm{E} = \omega_\mathrm{D} + \delta$, with $\delta \ll \omega_0$ (see Fig.~\ref{fig:freq}a).
Some frequency components of the force in Eq.~(\ref{eq:all_imps}) arise close to the cantilever resonance and therefore produce a deflection measurable with good SNR.
Numbering the IMPs by their position in the frequency domain with respect to~$\omega_\mathrm{D}$ (see Fig.~\ref{fig:imps}), we give the frequency~$\omega$ and the Fourier coefficients~$\hat{F}_\mathrm{TS}(\omega)$ in the table below:

\begin{center}
\begin{tabular}{| c | c | c |}
\hline
\textbf{IMP} & $\boldsymbol{\omega}$ & $\boldsymbol{\hat{F}_\mathrm{TS}(\omega)}$ \\
\hline
0 & $\omega_\mathrm{D}$ & $c_1 v_0 \mathrm{e}^{\mathrm{i}(\phi_1 + \theta_0)}$\\
\hline
1 & $\omega_\mathrm{E} = \omega_\mathrm{D} + \delta$ & $c_0 v_1 \mathrm{e}^{\mathrm{i}(\theta_1 - \phi_0)}$\\
2 & $2\omega_\mathrm{E} - \omega_\mathrm{D} = \omega_\mathrm{D} + 2\delta$ & $c_1 v_2 \mathrm{e}^{\mathrm{i}(\theta_2 - \phi_1)}$\\
3 & $3\omega_\mathrm{E} - 2\omega_\mathrm{D} = \omega_\mathrm{D} + 3\delta$ & $c_2 v_3 \mathrm{e}^{\mathrm{i}(\theta_3 - \phi_2)}$\\
\hline
-1 & $2\omega_\mathrm{D} - \omega_\mathrm{E} = \omega_\mathrm{D} - \delta$ & $c_2 v_1 \mathrm{e}^{\mathrm{i}(\phi_2 - \theta_1)}$\\
-2 & $3\omega_\mathrm{D} - 2\omega_\mathrm{E} = \omega_\mathrm{D} - 2\delta$ & $c_3 v_2 \mathrm{e}^{\mathrm{i}(\phi_3 - \theta_2)}$\\
-3 & $4\omega_\mathrm{D} - 3\omega_\mathrm{E} = \omega_\mathrm{D} - 3\delta$ & $c_4 v_3 \mathrm{e}^{\mathrm{i}(\phi_4 - \theta_3)}$\\
\hline
n & $(1-n)\omega_\mathrm{D} + n\omega_\mathrm{E} = \omega_\mathrm{D} + n\delta$ & $c_{1-n} v_n \mathrm{e}^{\mathrm{i}(\phi_{1-n} + \theta_n)}$\\
\hline
\end{tabular}
\end{center}
These Fourier coefficients of the force around resonance depend on both the capacitance gradient (coefficients $\{c_k, \phi_k\}$) and the electrical response (coefficients $\{v_j, \theta_j\}$).
However, we notice that the ratio and product of pairs depend only on the electrical response:

\begin{subequations}
\begin{align}
    \left|\frac{\hat{F}_{n+2}}{\hat{F}_{-n}}\right| & = \frac{v_{n+2}}{v_n},\\
    \mathrm{Arg}\left({\hat{F}_{n+2}\hat{F}_{-n}}\right) & = \theta_{n+2} - \theta_n,
\end{align}
    \label{eq:ratio_high}
\end{subequations}
where we used the short-hand notation $\hat{F}_n = \hat{F}_\mathrm{TS}(\omega_\mathrm{D} + n\delta)$.
Thus we eliminate the dependence on the capacitance gradient from our analysis.
No model for $\partial C / \partial z$ is required and thereby we significantly decrease the number of free parameters.
Furthermore, the ratio of force components is independent of the calibration constants $k$ and $\alpha$.
We require only $\omega_0$ and $Q$, which are directly measured with high accuracy, without relying on additional models and calibrations\cite{Borgani2017,Sader2016}.

One problem with the resonant scheme is that electrical excitation so close to resonance can generate a relatively large Fourier coefficient $\hat{F}_\mathrm{TS}(\omega_\mathrm{E})$ that results in large response at $\omega_\mathrm{E}$, therefore violating the assumption of Eq.~(\ref{eq:zt}) and significantly decreasing the accuracy of the reconstruction method.
To avoid this problem we introduce two schemes where the electrical excitation is placed well below or above resonance.

\subsection{Sub-resonant excitation}
This scheme is analogous to that used in intermodulation EFM\cite{Borgani2014}, which mechanically drives the cantilever close to its resonance frequency at $\omega_\mathrm{D} \approx \omega_0$, while the electrical excitation is at a much lower frequency $\omega_\mathrm{E} = \delta \ll \omega_\mathrm{D}$ (see Fig.~\ref{fig:freq}b).
The frequency components of the force in Eq.~(\ref{eq:all_imps}) for $k=1$ are those close to the cantilever resonance.
With the numbering convention of Fig.~\ref{fig:imps}, their Fourier coefficients are:

\begin{center}
\begin{tabular}{| c | c | c |}
\hline
\textbf{IMP} & $\boldsymbol{\omega}$ & $\boldsymbol{\hat{F}_\mathrm{TS}(\omega)}$ \\
\hline
0 & $\omega_\mathrm{D}$ & $c_1 v_0 \mathrm{e}^{\mathrm{i}(\phi_1 + \theta_0)}$\\
\hline
1 & $\omega_\mathrm{D} + \omega_\mathrm{E} = \omega_\mathrm{D} + \delta$ & $c_1 v_1 \mathrm{e}^{\mathrm{i}(\phi_1 + \theta_1)}$\\
2 & $\omega_\mathrm{D} + 2\omega_\mathrm{E} = \omega_\mathrm{D} + 2\delta$ & $c_1 v_2 \mathrm{e}^{\mathrm{i}(\phi_1 + \theta_2)}$\\
3 & $\omega_\mathrm{D} + 3\omega_\mathrm{E} = \omega_\mathrm{D} + 3\delta$ & $c_1 v_3 \mathrm{e}^{\mathrm{i}(\phi_1 + \theta_3)}$\\
\hline
-1 & $\omega_\mathrm{D} - \omega_\mathrm{E} = \omega_\mathrm{D} - \delta$ & $c_1 v_1 \mathrm{e}^{\mathrm{i}(\phi_1 - \theta_1)}$\\
-2 & $\omega_\mathrm{D} - 2\omega_\mathrm{E} = \omega_\mathrm{D} - 2\delta$ & $c_1 v_2 \mathrm{e}^{\mathrm{i}(\phi_1 - \theta_2)}$\\
-3 & $\omega_\mathrm{D} - 3\omega_\mathrm{E} = \omega_\mathrm{D} - 3\delta$ & $c_1 v_3 \mathrm{e}^{\mathrm{i}(\phi_1 - \theta_3)}$\\
\hline
n & $\omega_\mathrm{D} + n\omega_\mathrm{E} = \omega_\mathrm{D} + n\delta$ & $c_1 v_n \mathrm{e}^{\mathrm{i}(\phi_1 + \theta_n)}$\\
\hline
\end{tabular}
\end{center}
Dividing all the measured force components by the force component at $\omega_\mathrm{D}$ gives a complex coefficient depending on the electrical response only:
\begin{equation}
    \frac{\hat{F}_n}{\hat{F}_0} = \frac{v_n}{v_0}\mathrm{e}^{\mathrm{i}(\theta_n - \theta_0)}.
    \label{eq:ratio_low}
\end{equation}
All measured IMPs near resonance are proportional to $c_1$, as opposed to $c_{1-n}$ for the resonant scheme.
For a smooth capacitance gradient the coefficients $c_k$ quickly drop in magnitude as $k$ increases.
Therefore IMPs of the same order typically have higher magnitude (better SNR) when they are measured with the sub-resonant scheme, in comparison with resonant scheme.

\subsection{Super-resonant excitation}
We excite the electrical response at frequency $\omega_\mathrm{E} = 2\omega_\mathrm{D} + \delta$, close to the second harmonic of $\omega_\mathrm{D}$ (see Fig.~\ref{fig:freq}c).
With this scheme, $\hat{F}_\mathrm{TS}(\omega_\mathrm{E})$ and all its harmonics fall far from the cantilever resonance where the linear response function of Eq.~(\ref{eq:chi}) is very small.
This trick is similar to that used by Dicke in one of the first implementations of a lock-in amplifier\cite{Dicke1946}, where the modulation frequency at $30~\mathrm{Hz}$ was locked to half the power-line frequency, so that a spurious pickup at $60~\mathrm{Hz}$ and its harmonics would not affect the measurement.

Only the down-converted IMPs arise close to the resonance, with Fourier amplitudes:

\begin{center}
\begin{tabular}{| c | c | c |}
\hline
\textbf{IMP} & $\boldsymbol{\omega}$ & $\boldsymbol{\hat{F}_\mathrm{TS}(\omega)}$ \\
\hline
0 & $\omega_\mathrm{D}$ & $c_1 v_0 \mathrm{e}^{\mathrm{i}(\phi_1 + \theta_0)}$\\
\hline
1 & $\omega_\mathrm{E} - \omega_\mathrm{D} = \omega_\mathrm{D} + \delta$ & $c_1 v_1 \mathrm{e}^{\mathrm{i}(\theta_1 - \phi_1)}$\\
2 & $2\omega_\mathrm{E} - 3\omega_\mathrm{D} = \omega_\mathrm{D} + 2\delta$ & $c_3 v_2 \mathrm{e}^{\mathrm{i}(\theta_2 - \phi_3)}$\\
3 & $3\omega_\mathrm{E} - 5\omega_\mathrm{D} = \omega_\mathrm{D} + 3\delta$ & $c_5 v_3 \mathrm{e}^{\mathrm{i}(\theta_3 - \phi_5)}$\\
\hline
-1 & $3\omega_\mathrm{D} - \omega_\mathrm{E} = \omega_\mathrm{D} - \delta$ & $c_3 v_1 \mathrm{e}^{\mathrm{i}(\phi_3 - \theta_1)}$\\
-2 & $5\omega_\mathrm{D} - 2\omega_\mathrm{E} = \omega_\mathrm{D} - 2\delta$ & $c_5 v_2 \mathrm{e}^{\mathrm{i}(\phi_5 - \theta_2)}$\\
-3 & $7\omega_\mathrm{D} - 3\omega_\mathrm{E} = \omega_\mathrm{D} - 3\delta$ & $c_7 v_3 \mathrm{e}^{\mathrm{i}(\phi_7 - \theta_3)}$\\
\hline
n & $(1-2n)\omega_\mathrm{D} + n\omega_\mathrm{E} = \omega_\mathrm{D} + n\delta$ & $c_{1-2n} v_n \mathrm{e}^{\mathrm{i}(\phi_{1-2n} + \theta_n)}$\\
\hline
\end{tabular}
\end{center}
Similar to the previous two schemes, we take the ratio and product of pairs of force components to eliminate the dependence on the capacitance gradient:

\begin{subequations}
\begin{align}
    \left|\frac{\hat{F}_{n+1}}{\hat{F}_{-n}}\right| & = \frac{v_{n+1}}{v_n},\\
    \mathrm{Arg}\left({\hat{F}_{n+1}\hat{F}_{-n}}\right) & = \theta_{n+1} - \theta_n.
\end{align}
    \label{eq:ratio_superhigh}
\end{subequations}

\section{Time resolution}
\label{sec:res}
To estimate the achievable time resolution, we point out that intermodulation spectroscopy measures the change in amplitude and phase of the cantilever deflection~$\hat{d}$ from its free value~$\hat{d}_\mathrm{free}$ [see Eq.~(\ref{eq:Fts_from_d})].

For the sub-resonant scheme (Fig.~\ref{fig:res}b), multiple cantilever oscillations (exactly $n_\mathrm{D}=\omega_\mathrm{D} / \delta \approx Q$) probe every electrical excitation.
The time resolution~$\Delta t$ of the reconstructed electrical response is therefore limited by the period~$T_\mathrm{D}$ of a single cantilever-oscillation cycle:
\begin{equation}
    \Delta t_\mathrm{sub} = T_\mathrm{D} = \frac{2\pi}{\omega_\mathrm{D}} \approx \frac{2\pi}{\omega_0}.
\end{equation}
For a typical $300~\mathrm{kHz}$ cantilever we get $\Delta t_\mathrm{sub} \approx 3~\mathrm{\mu s}$.
The time resolution is inversely proportional to the resonance frequency, and it can be improved by using shorter and stiffer AFM probes with higher resonance frequency.

For the resonant  scheme (Fig.~\ref{fig:res}a), $n_\mathrm{D}$ cantilever oscillations are matched by $n_\mathrm{E}=n_\mathrm{D}+1$ electrical excitations.
This mismatch causes the delay between each oscillation and excitation to grow during the measurement window in multiples of a ``base delay'' $T_\mathrm{D}-T_\mathrm{E}$, where~$T_\mathrm{E}$ is the repetition period of the electrical excitation.
As for an optical pump-probe experiment, it is this base delay that sets the time resolution:

\begin{equation}
    \Delta t_\mathrm{res} = T_\mathrm{D}-T_\mathrm{E} = \frac{2\pi(\omega_\mathrm{E}-\omega_\mathrm{D})}{\omega_\mathrm{D}\omega_\mathrm{E}} \approx \frac{2\pi\delta}{\omega_0^2}.
    \label{eq:dt_res}
\end{equation}
For a typical $300~\mathrm{kHz}$ cantilever and a $2~\mathrm{ms}$ measurement time, we get $\Delta t_\mathrm{res} \approx 6~\mathrm{ns}$.
The time resolution is inversely proportional to the square of $\omega_0$, indicating that higher-frequency AFM probes would improve $\Delta t$ much more than for the sub-resonant scheme.
Moreover, we note how Eq.~(\ref{eq:dt_res}) depends on the experimental parameter~$\delta$: in principle we can achieve an arbitrarily short time resolution by decreasing the frequency spacing~$\delta$, and therefore increasing the measurement time $T_\mathrm{M}=2\pi/\delta$.
In practice, however, other experimental details limit the achievable time resolution, such as the frequency stability of the reference oscillation, the sharpness of the excitation pulse, and the slow drift of the cantilever linear response function~(\ref{eq:chi}) due to temperature fluctuations.

Similarly, for the super-resonant scheme (Fig.~\ref{fig:res}c) we find:
\begin{equation}
    \Delta t_\mathrm{super} = T_\mathrm{D}-2T_\mathrm{E} = \frac{2\pi(\omega_\mathrm{E}-2\omega_\mathrm{D})}{\omega_\mathrm{D}\omega_\mathrm{E}} \approx \frac{2\pi\delta}{2\omega_0^2}.
\end{equation}
The time resolution is improved by a factor of 2 with respect to the resonant case.

Our explanation of the achievable time resolution implicitly assumes that the entire cantilever deflection is measurable with good SNR.
In the frequency domain, all the IMPs of~$\hat{F}_\mathrm{TS}$ should be measurable above the noise level.
In practice, however, this is not possible and only a few IMPs are detected with appreciable SNR near the cantilever resonance.
If we obtain coefficients $v_n\mathrm{e}^{\mathrm{i}\theta_n}$ up to order N, we can use the inverse DFT to obtain $V^2(t)$ (see Fig.~\ref{fig:fit}), up to an offset and a scaling factor as all Fourier coefficients are in units of $v_0\mathrm{e}^{\mathrm{i}\theta_0}$.
The reconstructed signal has a time resolution of approximately
\begin{equation}
    \Delta t_\mathrm{iDFT} = \frac{2\pi}{N \omega_\mathrm{E}}.
\end{equation}
For the sub-resonant scheme we typically have $N=10$ and $\Delta t = 2\pi/(N \delta) \approx 200~\mathrm{\mu s}$.
For the resonant scheme typically $N=5$ and $\Delta t \approx 2\pi/(N \omega_0) \approx 0.7~\mathrm{\mu s}$.

Figure~\ref{fig:fit} demonstrates this limitation with a numerical simulation of the dynamics of the AFM cantilever in the sub-resonant scheme.
The simulation includes realistic thermal and detector noise contributions, and the actual potential of the surface (blue solid line) is a square wave with exponential rise and fall edges.
The green solid line shows the response calculated with the inverse DFT of the coefficients $v_n\mathrm{e}^{\mathrm{i}\theta_n}$, obtained from 10 IMPs in the simulated deflection spectrum.
The curve captures the general shape of the response, but oscillations are clearly visible and the rise and fall edges are not as sharp as in the actual response.

To overcome this practical limitation, we introduce a model for the electrical response of the material.
We need only a few Fourier coefficients of $\widehat{V^2}$ to accurately determine the parameters of the model, as shown by the orange dashed line in Fig.~\ref{fig:fit}.
By assuming a functional form for the response, we exploit correlations in the measured IMPs, pushing the time resolution down to the theoretical limit.

\section{Model-based reconstruction of electrical response}
\begin{figure}
    \centering
    \includegraphics[width=\columnwidth]{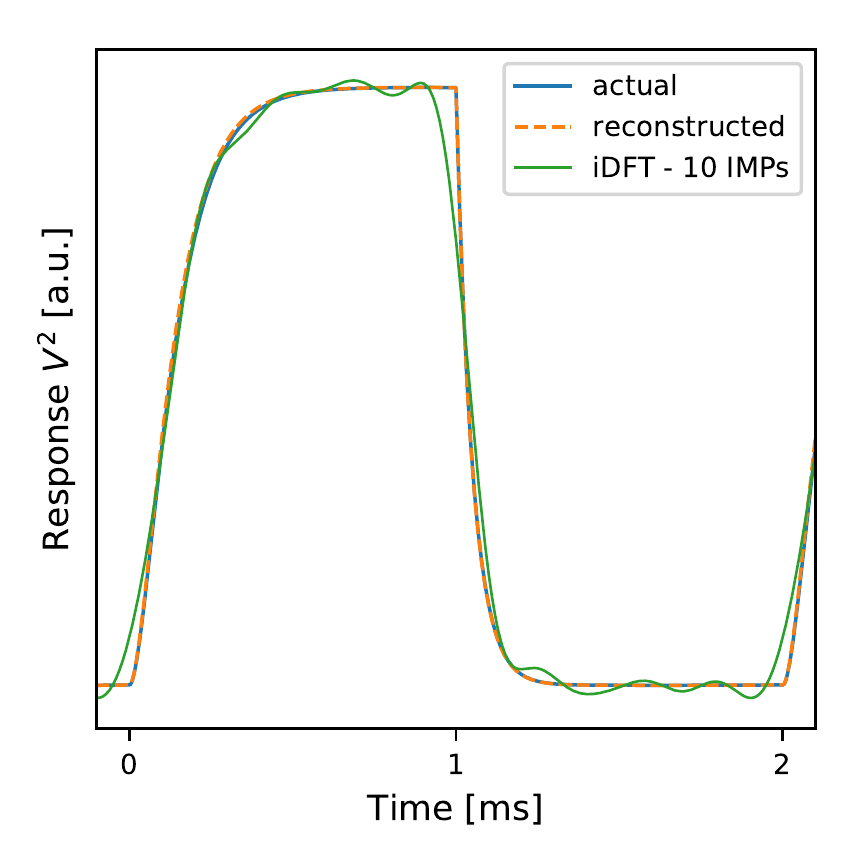}
    \caption{Square of the electrical response for a simulation of the sub-resonant scheme with noise.
    The actual response (solid blue) and the response reconstructed from the fit of the exponential model to the simulated cantilever dynamics (dashed orange) agree very well.
    The response calculated with the inverse DFT of 10 IMPs (solid green) follows the shape of the actual response, but oscillations are visible and the rise and fall slopes are smoothed.
    The actual response has $\tau_\mathrm{R}=\tau_\mathrm{F}=100~\mathrm{\mu s}$, the reconstructed response $\tau_\mathrm{R}=96.9~\mathrm{\mu s}$ and $\tau_\mathrm{F}=99.3~\mathrm{\mu s}$.
    }
    \label{fig:fit}
\end{figure}
\begin{figure}
    \centering
    \includegraphics[width=\columnwidth]{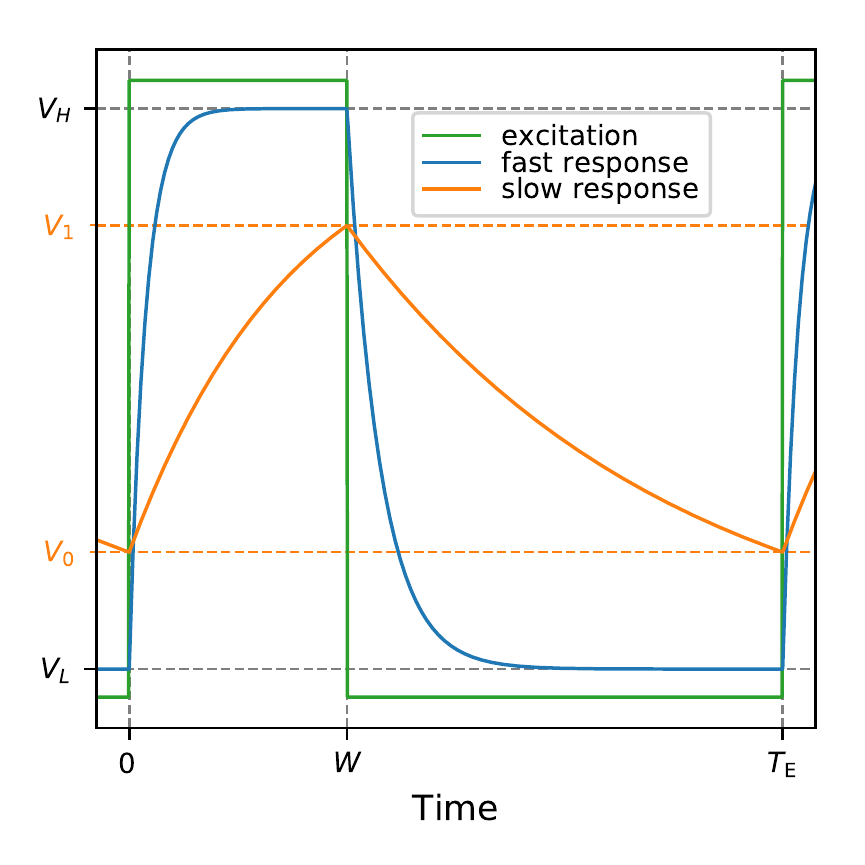}
    \caption{Model of the electrical response of the material (blue and orange) to a square-pulse excitation (green).
    $T_\mathrm{E}$ and $W$ are the pulse period and width, respectively.
    $V_\mathrm{H}$ and $V_\mathrm{L}$ are the equilibrium response when the excitation is high and low, respectively.
    $V_1$ and $V_0$ are the maximum and minimum response obtained due to the rise and fall time constants $\tau_\mathrm{R}$ and $\tau_\mathrm{F}$, respectively.
    For a fast response (blue), \textit{i.e.} $\tau_\mathrm{R} \ll W$ and $\tau_\mathrm{F} \ll T_\mathrm{E}-W$, we have $V_1 \approx V_\mathrm{H}$ and $V_0 \approx V_\mathrm{L}$.
    For a slow response (orange), we have $V_1 < V_\mathrm{H}$ and $V_0 > V_\mathrm{L}$.
    }
    \label{fig:pulse}
\end{figure}

We excite the sample with a periodic train of square pulses of length $W$ and repetition period $T_\mathrm{E}=2\pi/\omega_\mathrm{E}$ (green line in Fig.~\ref{fig:pulse}).
We model the electrical response of the sample as the blue and orange lines in Fig.~\ref{fig:pulse}, \textit{i.e.} with an exponential rise and exponential fall characterized by the time constants $\tau_\mathrm{R}$ and $\tau_\mathrm{F}$, respectively.
These time constants model material properties such as charge generation, diffusion and recombination.
$V_\mathrm{H}$ and $V_\mathrm{L}$ are the equilibrium response with and without excitation, respectively, modeling the change in contact potential difference due to the excitation.
In the case of fast dynamics, \textit{i.e.} small $\tau_\mathrm{R}$ and $\tau_\mathrm{F}$, the response reaches $V_\mathrm{H}$ and $V_\mathrm{L}$ within every pulse (blue line in Fig.~\ref{fig:pulse}).
However, for slow dynamics the response only reaches $V_1 < V_\mathrm{H}$ and $V_0 > V_\mathrm{L}$ (orange line in Fig.~\ref{fig:pulse}).
The highest and lowest values reached are
\begin{subequations}
\begin{align}
    V_1 & = \frac{(V_\mathrm{H}-V_\mathrm{L})(1 - \mathrm{e}^{-W/\tau_\mathrm{R}})}{1 - \mathrm{e}^{-W/\tau_\mathrm{R}-(T_\mathrm{E}-W)/\tau_\mathrm{F}}} + V_\mathrm{L},\\
    V_0 & = (V_1 - V_\mathrm{L}) \mathrm{e}^{-(T_\mathrm{E}-W)/\tau_\mathrm{F}} + V_\mathrm{L}.
\end{align}
\end{subequations}
The response $V$ as a function of the ``time-window coordinate'' $u = t \bmod T_\mathrm{E}$ is therefore
\begin{equation}
    V(u) = \begin{cases}
        (V_\mathrm{H} - V_0)(1-\mathrm{e}^{-u/\tau_\mathrm{R}}) + V_0 & \text{for $u < W$}\\
        (V_1 - V_\mathrm{L})\mathrm{e}^{-(u-W)/\tau_\mathrm{F}} + V_\mathrm{L} & \text{for $u > W$}.
    \end{cases}
\end{equation}

It is possible to calculate analytically the DFT of $V^2(t)$ at~$\omega=k\omega_\mathrm{E}$:
\begin{equation}
\begin{split}
    \widehat{V^2} &= \frac{\mathrm{e}^{-\mathrm{i}\omega\rho}}{T_\mathrm{E}} \left\{       V_\mathrm{H}^2 W ~ \operatorname{sinc}\left(W\frac{\omega}{2\pi}\right) \mathrm{e}^{-\mathrm{i} \omega W / 2}       \right.\\
    &+V_\mathrm{L}^2 \left(T_\mathrm{E} - W\right) ~ \operatorname{sinc}\left[\left(T_\mathrm{E} - W\right) \frac{\omega}{2\pi}\right] \mathrm{e}^{-\mathrm{i} \omega \frac{T_\mathrm{E} + W}{2}}\\
    &+ \left(V_\mathrm{H} - V_0\right)^2 \frac{\tau_\mathrm{R}}{2 + \mathrm{i} \omega \tau_\mathrm{R}} \left(1 - \mathrm{e}^{-\frac{2W}{\tau_\mathrm{R}} -\mathrm{i} \omega W}\right)\\
    &- 2 V_\mathrm{H} \left(V_\mathrm{H} - V_0\right)\frac{\tau_\mathrm{R}}{1 + \mathrm{i} \omega \tau_\mathrm{R}} \left(1 - \mathrm{e}^{-\frac{W}{\tau_\mathrm{R}} -\mathrm{i} \omega W}\right)\\
    &+ \frac{\left(V_1 - V_\mathrm{L}\right)^2\tau_\mathrm{F}}{2 + \mathrm{i} \omega \tau_\mathrm{F}} \left[1 - \mathrm{e}^{-\frac{2(T_\mathrm{E} - W)}{\tau_\mathrm{F}} -\mathrm{i} \omega (T_\mathrm{E} - W)}\right] \mathrm{e}^{-\mathrm{i} \omega W}\\
    &\left.       + \frac{2 V_\mathrm{L} \left(V_1 - V_\mathrm{L}\right) \tau_\mathrm{F}}{1 + \mathrm{i} \omega \tau_\mathrm{F}} \left[1 - \mathrm{e}^{-\frac{T_\mathrm{E} - W}{\tau_\mathrm{F}} -\mathrm{i} \omega (T_\mathrm{E} - W)}\right] \mathrm{e}^{-\mathrm{i} \omega W}       \right\}
    ,
    \label{eq:model}
\end{split}
\end{equation}
where~$\rho$ is a possible time delay in the electronics or measurement leads.
$\widehat{V^2}$ is thus a function of frequency depending on the known experimental parameters $T_\mathrm{E}$, $W$ and~$\rho$, and on four material parameters $V_\mathrm{H}$, $V_\mathrm{L}$, $\tau_\mathrm{R}$ and $\tau_\mathrm{F}$ to be determined.
We finally use a numerical least-square optimization routine to fit the material parameters to the measured Fourier coefficients of Eq.s~(\ref{eq:ratio_high}), (\ref{eq:ratio_low}) or~(\ref{eq:ratio_superhigh}), obtaining $V^2(t)$ (orange in Fig.~\ref{fig:fit}).

\section{Experimental results}
\begin{figure}
    \centering
    \includegraphics[width=\columnwidth]{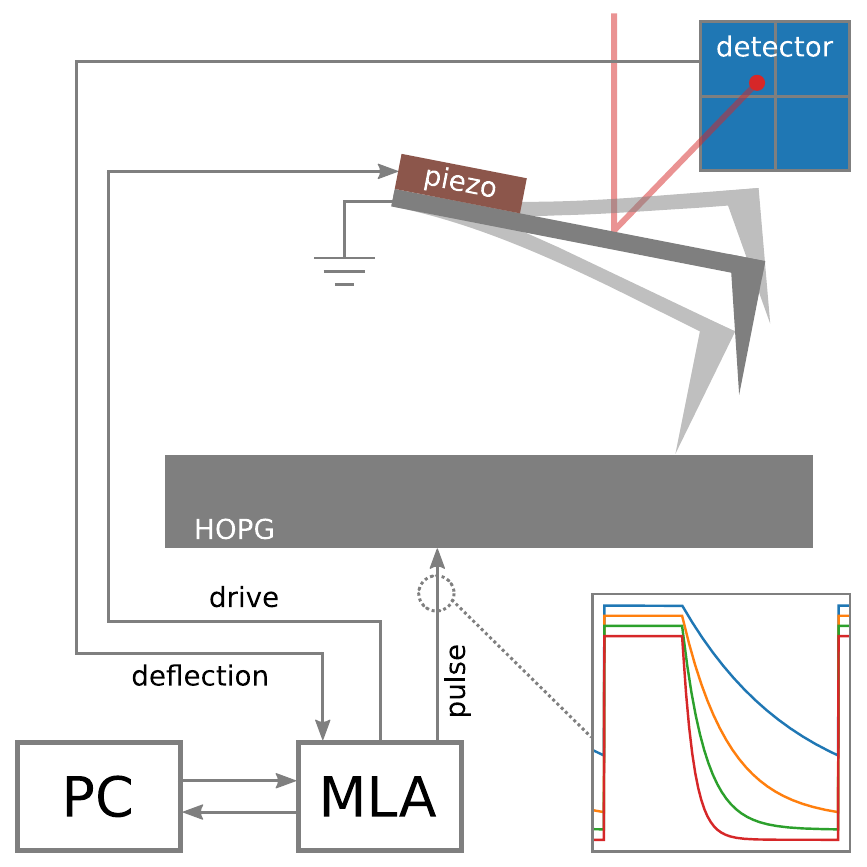}
    \caption{Schematic drawing of the measurement setup, not to scale.
    A multifrequency lock-in amplifier (MLA) is used to drive the cantilever with a piezoelectric shaker, to apply the electrical excitation to the sample, and to acquire the cantilever deflection from the optical-lever detection system.
    The conductive cantilever is kept at ground.
    A series of electrical pulses with different decay time are synthesized by the MLA and applied to a sample of highly oriented pyrolytic graphite (HOPG).
    The data from the MLA is sent to a personal computer (PC) for further analysis.
    }
    \label{fig:scheme}
\end{figure}
\begin{figure*}
    \centering
    \includegraphics[width=\textwidth]{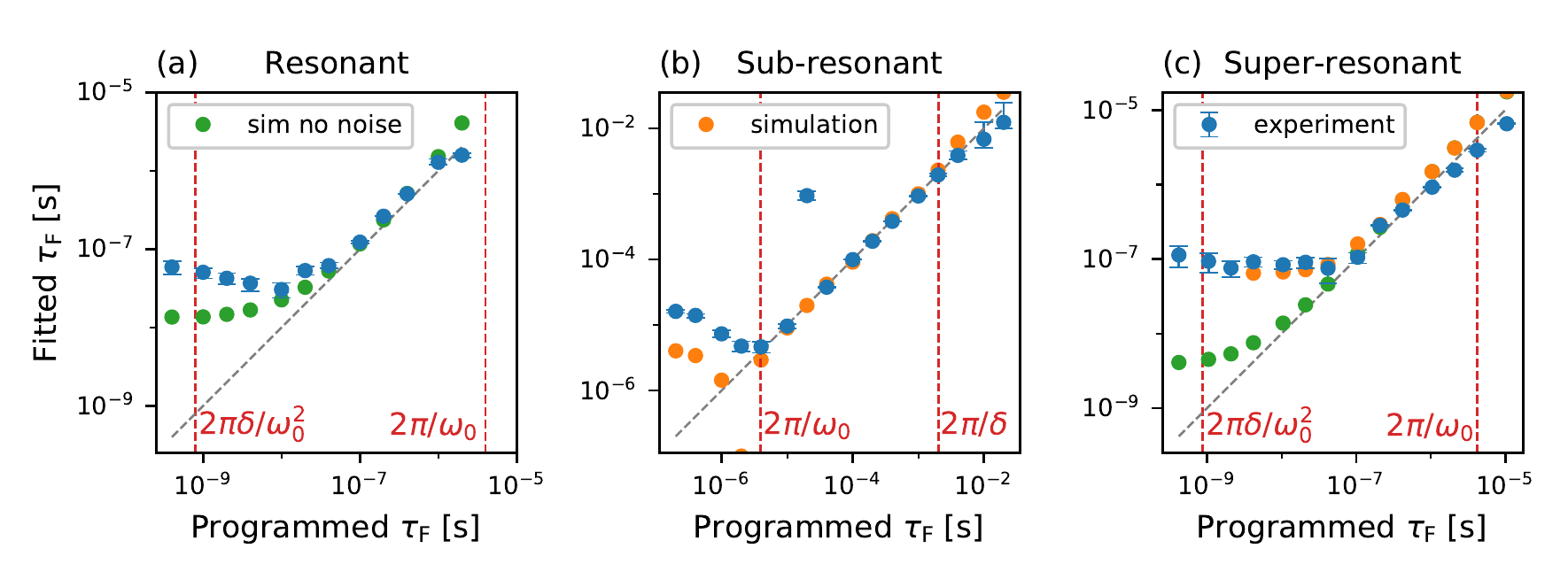}
    \caption{Results from simulations and from the experimental validation for the proposed excitation schemes: (a) resonant, (b) sub-resonant and (c) super-resonant.
    The cantilever has a resonance frequency of about $2\pi \times 250~\mathrm{kHz}$, a value of $\delta=2\pi \times 500~\mathrm{Hz}$ is used in the sub-resonant scheme and $\delta=2\pi \times 50~\mathrm{Hz}$ in the resonant and super-resonant schemes.
    The fitted values $\tau_\mathrm{F}$ are plotted versus the value programmed in the simulation and in the MLA.
    The gray dashed lines have slope unity and indicate where a perfect data point would be.
    For the experimental data, a series of 256 measurements is performed at each value of programmed $\tau_\mathrm{F}$: the blue dots indicate the median of the reconstructed values, and the error bars indicate the inter-quartile range.
    The vertical red dashed lines mark the time resolution calculated in Sec.~\ref{sec:res}.
    (a) in the resonant scheme, both simulations without noise (green dots) and experiments fail to reach the predicted time-resolution, due to the violation of Eq.~(\ref{eq:zt}).
    (b) in the sub-resonant scheme, simulations with detector and force noise (orange dots) and experiments show the predicted time resolution.
    (c) in the super-resonant scheme, simulations without noise approach the predicted time resolution, while experiments are limited to about $50~\mathrm{ns}$.
    Simulations with detector and force noise reproduce the experimental data.
    }
    \label{fig:proof}
\end{figure*}
Having verified the theoretical analysis with numerical simulation as described above, we next turn to experimental verification with programmed voltage pulses.
Figure~\ref{fig:scheme} is a schematic representation of the measurement setup.
A multifrequency lock-in amplifier\cite{IMP_AB} (MLA) drives the cantilever at resonance and monitors the amplitude and phase of the deflection at some 40 frequencies around resonance, sending the data to a computer where the fit to the analytical model of Eq.~(\ref{eq:model}) is performed to extract the material properties.
In addition, the MLA has a separate arbitrary-waveform functionality, with which we apply a programmed voltage pulse shape to a smooth and conductive sample (highly oriented pyrolytic graphite, HOPG).
Care was taken to transmit the pulses with $50~\mathrm{\Omega}$-matched impedance, as close as possible to the sample surface.
The electrical pulses affect the deflection of the AFM cantilever through the electrostatic tip-surface force.

In this control experiment, the electrical pulses are programmed with the shape shown in Fig.~\ref{fig:pulse}.
The pulse parameters $T_\mathrm{E}$, $W$, $V_\mathrm{H}-V_\mathrm{L}$ and $\tau_\mathrm{R}$ are known and fixed during the experiment.
The parameter $V_\mathrm{L}$ is \textit{a priori} unknown as it corresponds to the contact potential difference between the HOPG and the AFM tip.
Figure~\ref{fig:proof} shows the experimental data as well as the results of a numerical simulation.
The value of $\tau_\mathrm{F}$ obtained from the fitting routine is plotted versus the value programmed in the MLA.
For the sub-resonant scheme [Fig.~\ref{fig:proof}(b)], the fitted and programmed values agree above the time-resolution limit calculated in Sec.~\ref{sec:res}, in both simulated and experimental cases.

In the resonant scheme [Fig.~\ref{fig:proof}(a)], experiments and simulations do not reach the predicted time resolution and deviate from the ideal reconstruction for values of $\tau_\mathrm{F}$ below $20~\mathrm{ns}$.
No noise contribution is added to the simulated data, therefore we attribute this deviation to the violation of assumption in~(\ref{eq:zt}), as discussed above.

The super-resonant scheme approaches the theoretical time resolution in noise-free simulations [Fig.~\ref{fig:proof}(c)], but experimental results are limited to about $50~\mathrm{ns}$.
If we include both detector and force noise in the simulations, we can reproduce the experimental data, suggesting that the current time resolution is indeed limited by the noise in the measurement.
In the narrow band close to resonance used in our experiments, the limiting noise contribution is the thermal noise from Brownian motion of the AFM cantilever.
We therefore expect the method to perform better in vacuum conditions, where the increase in quality factor and decrease in thermal noise allows for greater force sensitivity.

We also observe an upper limit to the time resolution.
Even though the effect is not pronounced, the fitting of both the resonant (Fig.~\ref{fig:proof}a) and the super-resonant (Fig.~\ref{fig:proof}c) data starts to deviate from the programmed value of $\tau_\mathrm{F}$ when it exceeds $2\pi/\omega_0$.
Similarly, the fit gets worse in the sub-resonant case for values of $\tau_\mathrm{F}$ above $2\pi/\delta$.
In all three schemes, these upper limits correspond to $\tau_\mathrm{F}$ exceeding the period of the programmed voltage pulse $T_\mathrm{E}$, when the applied potential is effectively constant.
In these cases, a more traditional AFM technique like intermodulation electrostatic force microscopy\cite{Borgani2014} (ImEFM) or kelvin probe force microscopy\cite{Axt2018,Weber2018} (KPFM) that is able to measure the contact potential difference within  a time $2\pi/\delta\approx2~\mathrm{ms}$ would yield more accurate results.

\section{Conclusions}
We theoretically derived, simulated and demonstrated a novel AFM technique to measure the fast electrical dynamics of a material at the nanometer scale.
With the use of frequency multiplexing, only one measurement is required to obtain the time-evolution of a process, as opposed to a pump-probe scheme where multiple measurements for different pump-probe delays are necessary.
We achieved $\approx20~\mathrm{ns}$ time resolution with a commercially available AFM cantilever in ambient conditions, allowing for the mapping of nanosecond dynamics at standard tapping mode imaging speeds.

\begin{acknowledgments}
The authors acknowledge financial support from the Swedish Research Council (VR), and the Knut and Alice Wallenberg Foundation.
\end{acknowledgments}

\bibliography{my_bib}

\end{document}